# Bubble-assisted Liquid Hole Multipliers in LXe and LAr: towards "local dual-phase TPCs"

E. Erdal[1], L. Arazi*, A. Breskin, S. Shchemelinin, A. Roy*, A. Tesi, D. Vartsky, and S. Bressler

*Physics Faculty, Weizmann Institute of Science, Rehovot, Israel*

The bubble-assisted Liquid Hole Multiplier (LHM) is a novel concept for the combined detection of ionization electrons and scintillation photons in noble-liquid time projection chambers. It consists of a perforated electrode immersed in the noble liquid, with heating wires generating a stable vapor bubble underneath. Radiation-induced ionization electrons in the liquid drift into the electrode's holes and cross the liquid-vapor interface into the bubble where they induce electroluminescence (EL). The top surface of the electrode is optionally coated with a CsI photocathode; radiation-induced UV-scintillation photons extract photoelectrons that induce EL in a similar way. EL-photons recorded with an array of photo-sensors, e.g. SiPMs, provide event localization. We present the basic principles of the LHM concept and summarize the results obtained in LXe and LAr.

---

[1] Corresponding author; eran.erdal@weizmann.ac.il
*Currently at the Unit of Nuclear Engineering, Faculty of Engineering Sciences, Ben-Gurion University of the Negev, Beer Sheva, Israel

## Introduction

Noble-liquid detectors have become leading instruments in particle- and astroparticle physics; the main examples are Dark matter searches [1-9], Neutrino physics [10, 11] and rare event searches such as $\mu \rightarrow \gamma e$ $\mu \rightarrow e\gamma$ [12]. However, the scaling up of existing techniques, particularly that of dual-phase detectors, may encounter some technical difficulties - affecting the instruments' properties and thus the data quality. In light of this, we have proposed [13] and developed over the past few years a novel detection concept: the Liquid Hole Multiplier (LHM) [14-20]. The LHM, that can be considered as a "local dual-phase detector", permits detecting with a single element both ionization electrons and scintillation photons induced by radiation in the noble liquid.

The LHM concept is shown in Figure 1. A perforated electrode (e.g. a Gaseous Electron Multiplier (GEM) [21] or a Thick Gaseous Electron Multiplier (THGEM) [22]) is immersed in the noble liquid. Heating wires generate a vapor bubble, supported stably underneath the electrode. Radiation-induced ionization electrons in the liquid volume drift into the electrode's holes, cross the liquid-vapor interface and induce electroluminescence (EL) within the bubble (S2). The top surface of the electrode is optionally coated with a CsI photocathode; radiation-induced VUV-scintillation photons ("S1") extract photoelectrons that induce EL (S1') in a similar way. An array of photo-sensors located underneath the LHM electrode, measure the EL photons, providing a measurement of event's deposited energy and its 2D position. We expect that compared to a large-size "standard" dual-phase detector, where variations in the EL gap would affect energy resolution, the vapor bubble confined in a well-defined geometry and liquid-to-vapor interface, should yield superior results.

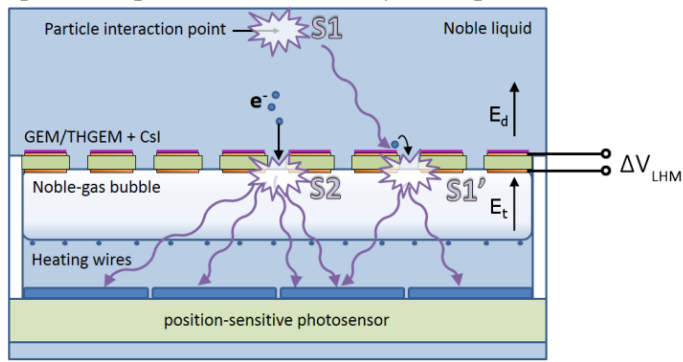

**Figure 1** Schematics of an LHM detector. A perforated electrode is immersed in the noble liquid, with a potential $\Delta V_{LHM}$ applied across. Radiation-induced ionization electrons and scintillation (S1)-induced photoelectrons from a CsI photocathode (deposited on top of the perforated electrode) are focused into the electrode's holes; both cross the liquid-vapor interface and create EL signals (S2 and S1', respectively) in the bubble.

The LHM concept has been studied extensively in a series of works, performed in dedicated cryostat systems (see details in [16] and in [20]), using small (~35 mm in diameter) prototypes. Most of the research has been performed in LXe [14-19] and recently in LAr [20]; the gases were recirculated and purified during the experiments.

## Summary of results

The response of the LHM detector has been investigated with $^{241}$Am sources immersed in the noble liquid. Four different types of electrodes were investigated in LXe: a THGEM with 0.3 mm in diameter holes, a "classical" GEM with bi-conical holes and two kinds of single-conical-hole GEMs (SC-GEM), differing in their thickness; the exact details are given in [18]. To date, the experiments in LAr have been performed only with a bare-THGEM electrode (detection of ionization electrons only). Experiments that included also primary-scintillation photon detection in LXe, comprised CsI photocathodes, deposited in a dedicated system [18].

EL-photon yields were measured with both PMTs and SiPMs, located underneath the perforated electrodes. In LXe, the PMT (Hamamatsu R8520) and the SiPM (Hamamatsu S13371) are both sensitive to the Xe emission line (175 nm). In LAr (128 nm), the PMT was vacuum-deposited with ~300 µg/cm² of full name of the 1,1,4,4-tetraphenyl-1,3-butadiene (TPB) [23] wavelength shifter; the SiPM had its protective quartz-window removed. Both photo-sensors permitted establishing the LHM response to ionization electrons and primary-scintillation photons, as function of the electrode configuration and various experimental parameters. The quad-SiPM also permitted reconstructing the shape of an annular [241]Am source [19, 20]. Signals were digitized and recorded using a Tektronix MSO5204B oscilloscope for post processing.

Examples of alpha-particle induced waveforms acquired in THGEM-based LHM detectors are shown in Figure 2; on the left, in LXe with a bare PMT and on the right, in LAr with a TPB-coated PMT. The LHM operated in LXe with a CsI photocathode on top of the THGEM electrode. Shown are pulses of a fraction of S1 primary-scintillation photons that passed through the electrode's holes to the PMT and two EL pulses: S1' due to scintillation-photoelectrons from CsI and S2 originating from the ionization electrons. The pulses on the right, with an LHM operating in LAr (without a photocathode), correspond to the S1 photons passing through the electrode's holes to the PMT and the ionization electrons generating an EL S2 pulse in the bubble.

One can clearly observe the difference in the pulse shape of the S2 signals in LXe and in LAr. In LXe, the pulse width can be explained by track of the electrons from the moment they cross the liquid-gas interface to the moment they reach the bottom electrode [18]. The long tail of the pulse in the LAr-LHM corresponds to the decay time of triplet states in Ar gas [24].

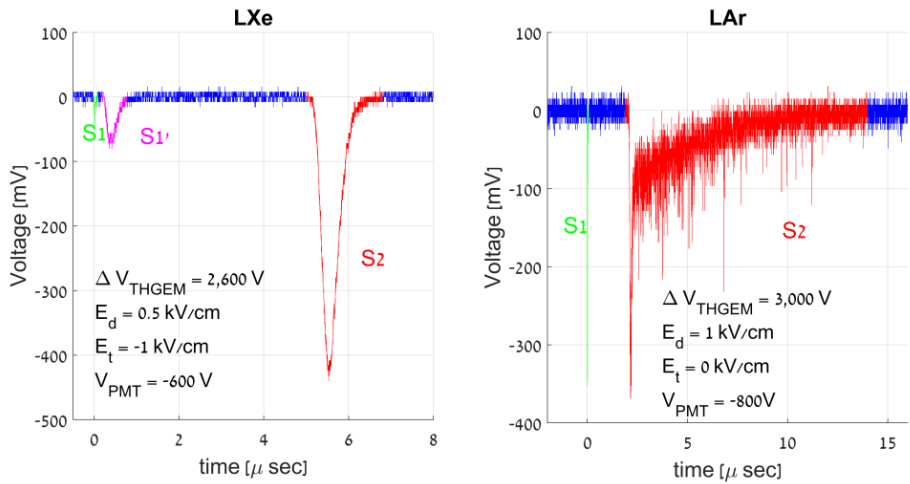

**Figure 2** Sample EL waveforms induced by alpha particles, recorded from a LHM detector: (left) in LXe and bare-PMT, with a CsI-coated THGEM electrode and (right) in LAr and TPB-coated PMT. Adapted from [18] and [20].

For each recorded waveform, the area under the curve was numerically computed and a histogram of the pulse areas was plotted. Figure 3 (left) shows an S2 (ionization electrons) energy spectrum induced by alpha particles in LXe, with a SC-GEM-LHM; Figure 3 (right) shows the energy spectrum of the S1' (scintillation photons) pulse in the same detector. A Gaussian fit was applied to the spectra to extract the mean and the resolution values. Similar histograms were also obtained in LAr with a two-fold worse resolution than in LXe [20], the difference in performance between the two liquids is under ongoing investigation.

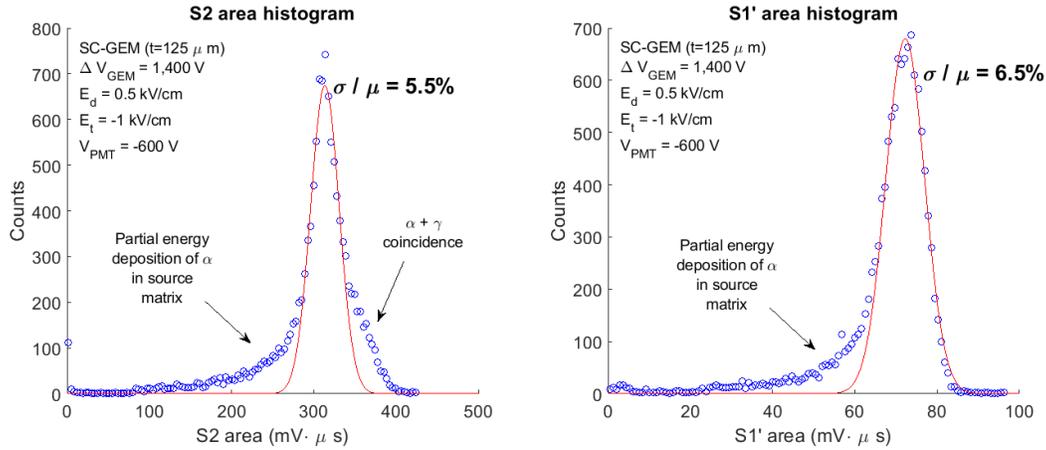

**Figure 3** Pulse-area EL spectra induced by alpha particles, recorded in LXe with a SC-GEM-LHM (left) ionization-electrons S2 spectrum; (right) S1' scintillation photons spectrum. From [18].

The mean values derived from the S2 pulse-area distributions are plotted in Figure 4 as a function of the voltage across the LHM electrode. Figure 4 (left) presents the S2 dependence on the voltage across the different electrodes investigated in LXe; Figure 4 (right) shows the dependence of S2 on the voltage across the THGEM electrode in LAr. While in LAr the data were not yet normalized, the highest EL light yield in LXe (i.e. the number of photons emitted from the LHM per drifting electron) was estimated to be ~400 photons/$e^-$/4$\pi$, with the 125 µm thick SC-GEM (see [18] for details).

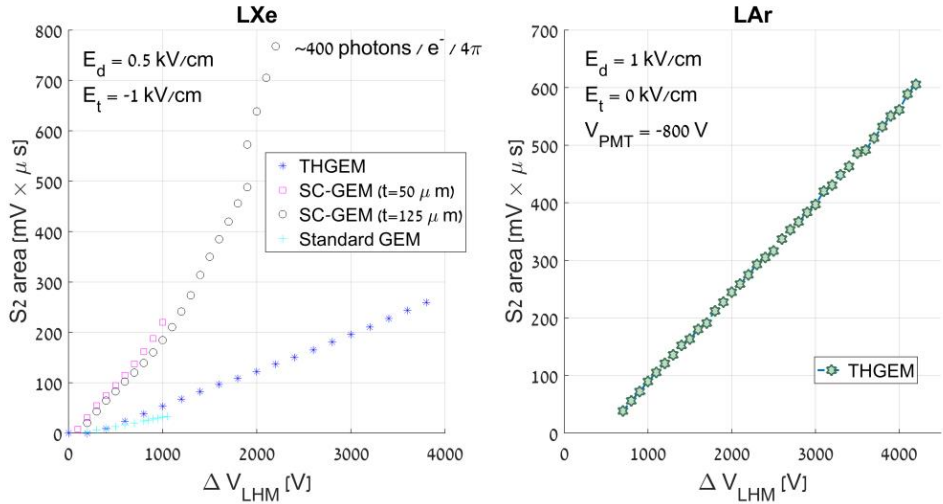

**Figure 4** S2 response of an LHM as a function of the voltage applied to the electrodes. The figure in LXe was adapted from [18] and the one in LAr is from [20].

In the measurements presented in Figure 4, performed with $E_t$=0, the dipole field at the hole vicinity directs electrons emitted into the bubble, to the bottom face of the electrode. Under these conditions, EL production is limited to this region. Applying an intense electric field, $E_t$, between the bottom part of the electrode and the heating wires (Figure 1) causes the electrons to generate EL photons also along their path towards the heating wires, with an increased emission at the high fields near the wires. Further increase results in moderate charge multiplication near the wires, followed by an increase in the photon yields [18]. An example of this process is reflected in the varying EL-pulse shapes presented in Figure 5(left), recorded in LXe for different $E_t$ values. Similar results were observed in LAr [20].

Figure 5 (right) presents the mean S2 pulse-area as a function of $E_t$ for a fixed LHM-electrode voltage. In addition to a rise in the EL-photon yield, the exponential trend suggests modest charge gain.

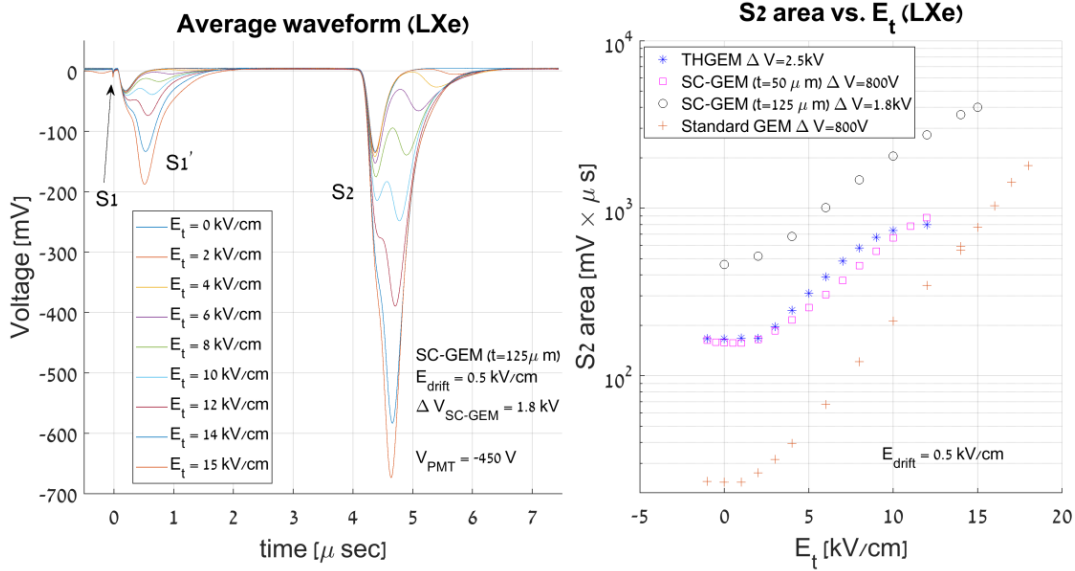

**Figure 5** (left) Average pulse shape in a SC-GEM LHM in LXe, at fixed LHM-electrode voltage and different values of $E_t$. (right) dependence of S2 pulse area on $E_t$.

Event localization was performed with a SiPM array located under the wires. Image reconstruction was done using a simple center of gravity method [19]. Figure 6 presents the auto-radiographic image of the 3.9 mm-diameter annular $^{241}$Am source and its reconstructed images in LXe [19] and in LAr [20] THGEM-LHM detectors; the data in LAr was taken under non-optimal conditions and therefore we expect it to improve in future works [20].

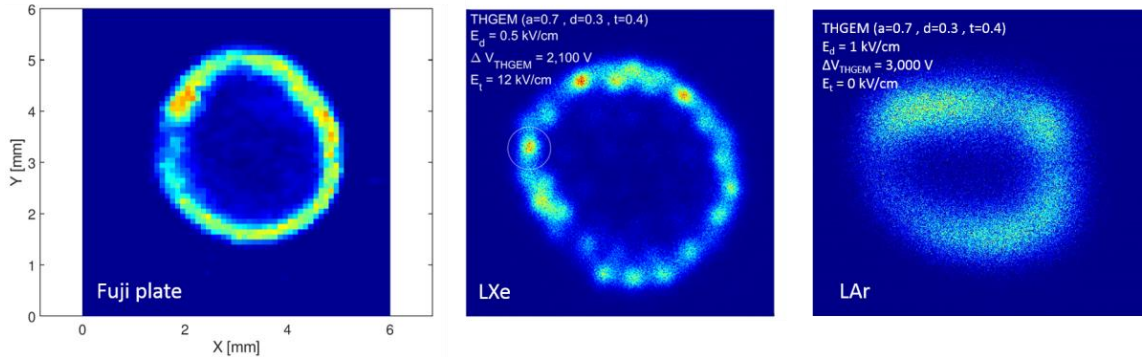

**Figure 6** (left) Auto-radiographic image of the annular $^{241}$Am source and its reconstructed images in a THGEM-LHM read out with a quad-SiPM sensor in (middle) in LXe and (right) in LAr. Note the reconstructed holes of the THGEM electrode in LXe.

As described above, a CsI-coated LHM electrode acts also as a UV-photon detector, with a photon detection efficiency (PDE) value dictated by the photocathode's vacuum quantum efficiency (QE) at a given wavelength, photoelectron extraction efficiency into the liquid (the multiplication of both provides the effective QE ($QE_{eff}$)), their collection efficiency into the holes and transmission efficiency into the bubble. The vacuum QE values of CsI are ~23% at Xe wavelength [25] and ~70% at that of Ar wavelength [26]. A first method involved estimating the expected PDE based on the QE$_{eff}$ of CsI immersed in LXe. As can be seen in Figure 7(left), following normalized photocurrent measurements, already at rather modest fields (above ~3 kV/cm), the photocurrent corresponds to QE$_{eff}$~20%; it reaches ~30% above 10 kV/cm. Similar results were also reported in [27]. When computing the electric field map on the surface of an electrode (e.g. a standard GEM as in Figure 7(right)), one can average the QE$_{eff}$ across the surface of the electrode to estimate the overall expected QE$_{eff}$. The resulting average expected value in a GEM or a SC-GEM (biased almost to their respective maximal voltages before sparking) is QE$_{eff}$~20%. The low PDE value of 3-4% measured in LXe in a SC-GEM-LHM [18] suggests a combined low photoelectron collection efficiency into the holes and their transmission efficiency into the bubble. The physics processes at the origin of this problem are discussed in [18] and are part of current studies in LXe and LAr.

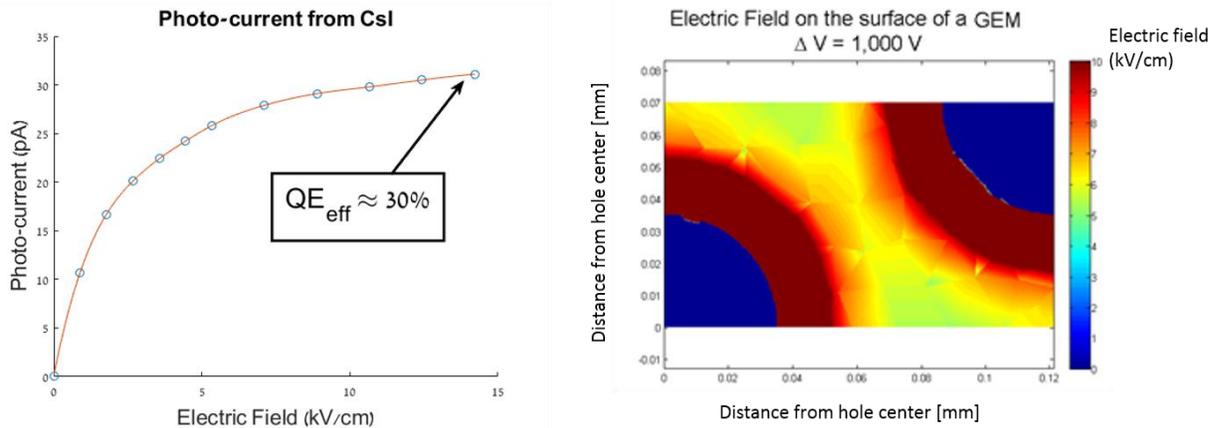

Figure 7 (left) Measured photocurrent and estimated Effective QE of CsI photocathode immersed in LXe, illuminated with LXe scintillation photons. (right) COMSOL simulation of the electric field on the surface of a unit cell of a GEM electrode.

**Discussion**

The new noble-liquid LHM concept, relying on EL in a stable vapor bubble formed under a hole-electrode immersed in the liquid, may evolve in the future into a viable charge and light sensor for large-volume radiation detectors. The concept has been validated with small prototypes in LXe and lately in LAr (in the latter, so far, for ionization-electron detection).

While showing promising results, the data presented here raise interesting questions, calling for deeper investigations of the basic underlying physics processes governing the LHM operation. While the basic interaction processes in noble liquids are well understood, the considerable dependence of the EL-photon yields measured in LXe with different electrode geometries is not yet well understood. Similarly, the unexpectedly low PDE values. In both cases, the origin could be related to the collection efficiency of electrons (from the liquid) and that of photoelectrons (from CsI) into the holes. In addition, the location and shape of the liquid-to-gas interface at the hole's bottom may affect the electron transfer efficiency into the gas bubble. As discussed in [18], some of the electrons reaching the interface do not necessarily cross it, due to the potential barrier; they may rather 'glide' sideways, towards the hole's bottom circumference - without generating EL. Both effects are the subject of current investigations; so is the two-fold lower energy resolution compared to that of LXe.

A serious improvement of the PDE value of the LXe-LHM (e.g. >15%) would permit conceiving large-volume local dual-phase LHM-based TPCs, e.g. for future dark-matter searches. A potential scheme, proposed within the DARWIN experiment R&D program [4] is shown in Figure 8. It consists of covering the TPC bottom with LHM modules, of the kind shown in Figure 1.

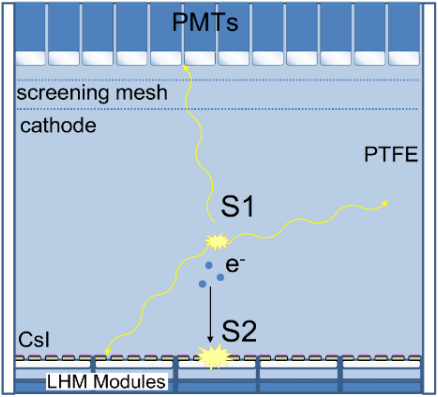

**Figure 8** Schematic design of a liquid-only single-phase TPC with bubble-assisted LHM modules at the bottom. They detect S1 signals of a fraction of the scintillation photons and S2 signals created by electrons drifting down towards the LHM. Accurate S2-based position reconstruction is provided by an array of photosensors below the LHMs.

**Acknowledgements**
We would like to thank Dr. M. L. Rappaport (Weizmann Institute of Science - WIS) his invaluable advices with the cryogenic systems' design. We also thank Dr. N. Canci (INFN-LNGS), assisted by M. Weiss (WIS) for the TPB deposition. This work was partly supported by the Israel Science Foundation (Grant No. 791/15) and by the I-CORE Program of the Planning and Budgeting Committee. The research is part of the DARWIN generic R&D program and within CERN-RD51 detector R&D collaboration. This work is supported by Sir Charles Clore Prize. Special thanks to Martin Kushner Schnur for supporting this research.